\documentclass{article}
\usepackage[dvips]{graphics}

\begin{document}

\author{K. Splittorff\thanks{email: split@alf.nbi.dk}\hspace{2mm} and 
A.D. Jackson\thanks{email: jackson@alf.nbi.dk}\\ 
The Niels Bohr Institute \\ 
Blegdamsvej 17\\ 
DK-2100 Copenhagen \O 
\\ Denmark}
\title{The Ginsparg-Wilson relation and local chiral random matrix theory}

\maketitle
\begin{abstract}
A chiral random matrix model with locality is constructed and examined. 
The Nielsen-Ninomiya no-go theorem is circumvented by the use of a generally 
applicable modified Dirac operator which respects the Ginsparg-Wilson 
relation.  We observe the expected universal behaviour of the eigenvalue 
density in the microscopic limit.
\end{abstract}
\vspace{3mm}

{\bf \large Introduction}
\vspace{3mm}
 
The success of chiral random matrix models, $\chi$RMM, in describing 
certain aspects of non-pertubative QCD indicates that the correlators 
in the Dirac spectrum are universal and suggests that the nature of 
the spontaneous breaking of chiral symmetry does not depend on the 
detailed dynamical properties of the gauge field. (See Jackson and 
Verbaarschot \cite{VEERJACK1} and Wettig, Sch\"afer, and 
H.A. Weidenm\"uller \cite{WSW}.  For a comprehensive review, see 
Verbaarschot \cite{VERrew}.)  Here, we want to extend these considerations 
by constructing a local $\chi$RMM through the explicit inclusion of 
the Euclidean derivative terms of the free Dirac operator.  There are 
several reasons for the construction of such models.  In practice, the 
range of validity of random matrix theory in describing real lattice 
gauge simulations is determined by the ability of the gauge field to mix 
free quark eigenstates.  (In condensed matter physics, this range is 
known as the ``Thouless energy''.)  The completely democratic treatment 
of all basis states in usual chiral random matrix models makes it impossible 
to address this question.  For the same reason, local quantities, e.g., 
propagators, cannot be considered in usual $\chi$RMM.  This limitation 
can be overcome by the construction of a local $\chi$RMM.

In order to represent the Euclidean derivative by a finite size matrix, it is 
necessary to discretize a finite (Euclidean) space-time volume.  This leads 
inevitably to the problem of unintended fermion doubling.  If one insists 
that chiral symmetry maintain its canonical form, $\{D,\gamma_5\}=0$, on 
the lattice, this doubling cannot be avoided.  In order to obtain 
a sensible local $\chi$RMM, we must resolve the fermion doubling in a 
manner which preserves a natural extension of chiral symmetry for finite 
volumes and lattice spacings.  The impossibility of constructing a lattice 
Dirac operator, $D$, with 
locality and exact chiral symmetry but free of fermion doubling is a 
consequence of the Nielsen-Ninomiya no-go theorem \cite{NandN}.  Some 
years ago, Ginsparg and Wilson \cite{GW} suggested the possibility of 
circumventing this no-go theorem by modifying the chiral symmetry condition to 
the form $\{ D , \gamma_5 \} = a D \gamma_5D$.  Following the recognition by 
 Hasenfratz \cite{Harse0} that fixed point actions of QCD satisfy the 
Ginsparg-Wilson relation, attention has again been drawn to lattice 
theories which satisfy the Ginsparg-Wilson relation, see Hasenfratz, Laliena,
 and Niedermayer \cite{HarseI}, Hasenfratz \cite{HarseII}, Neuberger 
\cite{neuI,neuII}, L\"uscher \cite{luscherII}, and Narayanan \cite{rajaI}. 
 L\"uscher \cite{luscherII} has shown that the Ginsparg-Wilson relation 
ensures an exact ``lattice chiral symmetry'' of the fermion action for any 
finite lattice size.  Because of this symmetry and the fact that the explicit 
breaking of chiral symmetry enters through a simple condition on $D$, the 
Ginsparg-Wilson approach suggests itself as a suitable way of avoiding the 
no-go theorem in a local $\chi$RMM.  Adopting this approach, it is possible 
to find a resolution of the fermion doubling problem and embed it in a 
local $\chi$RMM. 

The object of the present letter is to present a general procedure for the 
construction of a Dirac operator which satisfies the Ginsparg-Wilson relation 
and solves the fermion doubling problem.  This procedure is applicable to 
lattice QCD as well as to the local $\chi$RMM which we construct.  Further, we 
shall consider the spectral properties of this local $\chi$RMM and 
demonstrate the universality of the microscopic spectral density, i.e., 
the spectral density near eigenvalue zero on the scale of the lowest 
Dirac eigenvalue.
\vspace{3mm}

{\bf \large \noindent The Ginsparg-Wilson relation and its implementation}
\vspace{3mm}

The Ginsparg-Wilson (GW) relation is a constraint on the Dirac 
operator~$D$
\begin{equation}
D\gamma_5+\gamma_5D=aD\gamma_5D \ ,
\label{GWrel}
\end{equation} 
where $a$ is the lattice spacing.  The Dirac operator appears as the heart 
of the fermionic sandwich, $\bar{\psi}D\psi$, in the fermionic part of the 
QCD action 
\begin{equation}
{\cal S}_{F}=a^4\sum_x \bar{\psi}D\psi \ .
\label{AQCD}
\end{equation}  
Since the anti-commutator of $D$ and $\gamma_5$ is not zero, chiral 
symmetry is explicitly broken.  Instead, as shown by L\"uscher 
\cite{luscherII}, a new ``lattice chiral symmetry'' is present.  The 
infinitesimal variation of the fermion fields associated with 
this new symmetry is 
\begin{eqnarray}
\psi  \to  \psi+\varepsilon\gamma_5(1-\frac{1}{2}aD)\psi & \mbox{and} &
\bar{\psi} \to \bar{\psi}+\varepsilon\bar{\psi}(1-\frac{1}{2}aD)\gamma_5 \ .
\label{infsym} 
\end{eqnarray} 
Extension of this singlet flavour lattice chiral symmetry to a non-singlet 
lattice chiral symmetry is straightforward.  Both variations are symmetries 
of the fermionic action provided that the Dirac operator satisfies the GW 
relation.  Thus, the GW relation ensures the presence of a continuous 
symmetry which can be regarded as the remnant of chiral symmetry on finite 
lattices.  The Nielsen-Ninomiya no-go theorem states that lattice Dirac 
operator cannot simultaneously satisfy a series of physically reasonable
demands.  Among these is that $\{D,\gamma_5\}=0$.  By explicitly modifying 
the form of chiral symmetry through the GW relation to the form of 
eqn.\,(\ref{infsym}), the no-go theorem can be circumvented.  It is 
necessary to adopt a strategy for dealing with the fermion doubling 
problem.   Here, we choose to work with the Wilson lattice realization 
of the continuum operator, $H=\gamma_5[1-\gamma_\mu(\partial_\mu+iA_\mu)]$, 
but this choice is not mandatory.  Let us first see how to implement 
the GW relation. 

Consider the Dirac operator, $D$, in an arbitrary matrix representation.  
Since $D$ is to be multiplied by the row, $\bar{\psi}$, from the left and 
by the column, $\psi$, from the right, $D$ must be a square matrix of 
dimension, say, $2N\times 2N$.  For some $2N \times 2N$ matrix $\epsilon(H)$, 
we can write $D$ as
\begin{equation}
D=\frac{1}{a}({\bf 1} -\gamma_5\epsilon(H)) \ ,
\label{defD}
\end{equation}
where\footnote{Hasenfratz \cite{HarseI} has shown that the index theorem is 
reproduced for finite lattice spacing and size provided that the GW relation 
holds.  Thus, the difference in the number of $1$ and $-1$ entries in 
$\gamma_5$ depends on the topological index, $\nu=n_+ - n_-$, which we 
shall take as zero.} $\gamma_5={\rm diag}(1,1,..,1,-1,-1,..,-1)$.  
Since $\gamma_5$ is invertible, there is a one-to-one correspondence between 
$D$ and $\epsilon(H)$.  As noted by Narayanan \cite{rajaI}, the requirement 
that $D$ satisfies the GW relation is equivalent to 
\begin{equation}
(\epsilon(H))^2={\bf 1} \ .
\label{GWRepsilon}
\end{equation}  
Since the dynamics of $D$ enters through $\epsilon(H)$, one must 
build into $\epsilon(H)$ some regularization of the na\"{\i}ve 
discretization which effectively eliminates fermion doubling.  Having 
chosen a suitable form for $H$, we define the matrix $\epsilon(H)$ as
\begin{equation}
\epsilon(H) \equiv U_H \, {\rm diag}({\rm sign}(h_1),{\rm sign}(h_2),..,
{\rm sign}(h_{2N})) \, U^\dagger_H
\label{defepsilon}
\end{equation}
where $U_H$ is the unitary matrix that diagonalizes $H$,  
\begin{equation}
U^\dagger_H H U_H={\rm diag}(h_1,h_2,..,h_{2N}) \ .
\end{equation}
While any choice of $H$ which provides a suitable regularization of the 
na\"{\i}ve discretization can be adopted, it is essential for the 
construction of $\epsilon(H)$ that $H$ be hermitian (or antihermitian) 
and that ${\rm \det}(H)\not=0$.
\vspace{2mm}

Before turning to the construction of a local chiral random matrix model, 
it is useful to consider some general properties of the spectrum of the 
resulting Dirac operator and of the order parameter for the finite lattice 
chiral symmetry.  Since $\epsilon(H)$ is hermitian and satisfies 
eqn.\,(\ref{GWRepsilon}), the combination $\gamma_5\epsilon(H)$ is 
unitary.  This implies that $D$ is normal and that the spectrum of $D$ lies 
in the complex plane on the circle, $(1-e^{i\theta})/a$ with 
$\theta\in[-\pi,\pi]$.  Furthermore, $D$ satisfies the hermiticity relation, 
$\gamma_5D\gamma_5=D^\dagger$.  As noted in \cite{FLW}, this property in 
conjunction with the GW relation implies a complex conjugation symmetry in the 
spectrum of $D$: 
\begin{eqnarray}
D\psi_\lambda & = & \lambda\psi_\lambda \\
D\gamma_5\psi_\lambda & = & \lambda^*\gamma_5\psi_\lambda \mbox{\hspace{4mm} 
if $\lambda^*\not=\lambda$} \\
\gamma_5\psi_\lambda & = & \pm\psi_\lambda  \mbox{\hspace{8mm} 
if $\lambda^*=\lambda$} \ . 
\end{eqnarray}
Using this complex conjugation symmetry, a Banks-Casher-like relation 
\cite{BC} appears in the limits ${\rm volume}\to\infty$ then $m\to 0$, where 
$m$ is a regulator mass.
\begin{eqnarray}
\langle\bar{\psi}\psi\rangle & = & \langle D^{-1}(x,x)\rangle\label{average}\\
& = & \lim_{m \to 0}\lim_{{\rm Vol} \to \infty}\lim_{a \to 0} 
\frac{1}{{\rm Vol}}\int_{-\pi/a}^{\pi/a}
\frac{\rho(s)}{(1-e^{as})/a+m}ds\label{BCudregning}\\
 & = & \lim_{m\to0}\frac{1}{{\rm Vol}} 
\int_0^\infty\frac{2m\rho(s)}{s^2+m^2}ds \ ,
\end{eqnarray}
where the last equality relies on the identity $\rho(s)=\rho(-s)$, which is 
a consequence of the complex conjugation symmetry. The average in 
eqn.\,(\ref{average}) is over gauge field configurations.  Finally, we 
obtain the Banks-Casher relation for the order parameter in the limit of 
zero lattice spacing 
\begin{equation}
\langle\bar{\psi}\psi\rangle=\frac{\pi\rho(0)}{{\rm Vol}} \ .
\label{thetaBC}
\end{equation}

It is also straightforward to evaluate the integral in 
eqn.\,(\ref{BCudregning}) in the limit $m\to0$ for any finite $a$.  
One obtains
\begin{equation}
\frac{a}{{\rm Vol}}\int^{\frac{\pi}{a}}_0\rho(s)ds+ 
\frac{\pi\rho(0)}{{\rm Vol}} \ .
\label{finiteBC}
\end{equation}
Hasenfratz \cite{HarseII} has suggested the ``subtracted'' chiral condensate 
as a possible order parameter for lattice chiral symmetry,
\begin{equation}
\label{16}
\langle\bar{\psi}\psi\rangle_{{\rm sub}}=\frac{1}{\rm Vol} 
\langle{\rm Tr}(D^{-1}-\frac{1}{2}a{\bf 1})\rangle \ , 
\end{equation}
where the trace extends over colour, flavour, and Dirac space indices.  
By virtue of the GW relation, only zero modes of $D$ contribute to 
$\langle\bar{\psi}\psi\rangle_{{\rm sub}}$.  The result of 
eqn.\,(\ref{finiteBC}) is that a Banks-Casher relation remains valid 
at all finite lattice spacings provided one uses the subtracted chiral 
condensate as the order parameter for the lattice chiral symmetry.  
Of course, in the limits of infinite volume and $a \to 0$, the lattice 
chiral symmetry becomes genuine chiral symmetry, and the subtracted 
chiral condensate evolves into the true chiral condensate.
\newpage

{\bf \large \noindent A local chiral random matrix model}
\vspace{3mm}

We shall construct a local $\chi$RMM using the standard Wilson prescription 
for dealing with the fermion doubling problem.  We choose as a basis the 
states obtained by chiral projection from the eigenstates of the na\"{\i}ve 
lattice realization of the free Euclidian continuum operator $\gamma_\mu 
\partial_\mu$.  In this basis, $H$ is defined as
\begin{equation}
H\equiv\gamma_5X ,\mbox{ where }X\equiv {\bf 1}
-a\left[\left(\begin{array}{cc}
      \Delta' & i\Delta \\ 
      i\Delta & \Delta' 
    \end{array}\right)
+\left(\begin{array}{cc}
      0 & iW \\ 
      iW^\dagger & 0 
    \end{array}\right)\right] \ .
\label{Hmodel}
\end{equation}
In general, the matrix $W$ is uniquely determined by the gauge field 
configuration.  The local $\chi$RMM Dirac operator is defined through 
eqns.\,(\ref{defD}) and (\ref{defepsilon}) by replacing $W$ by a 
suitable random matrix.  The $N\times N$ matrices $\Delta$ and $\Delta'$ 
are real and diagonal with 
\begin{eqnarray}
\Delta & = & \frac{1}{a}{\rm diag}(\ldots ,[\sum_{i=1}^3 
\sin^2(ap^{(j)}_i)+\sin^2(a\pi T)]^{\frac{1}{2}}, \ldots )\label{defdelta}
\\
\Delta' & = & \frac{1}{a}{\rm diag}(\ldots ,\sum_{i=1}^3 
(1-\cos(ap^{(j)}_i))+ 1-\cos(a\pi T), \ldots ) \ ,
\label{defdeltap}
\end{eqnarray}
where $-n/2 \leq j \leq n/2-1$ with $n \equiv N^{1/3}$.  The 
spatial momenta \\ $p^{(j)}_i=2\pi j/(an)$ are determined by imposing 
periodic boundary 
conditions in the spatial directions.  Due to the anti-periodic 
boundary conditions in the time direction, the temporal momenta 
are given by the Matsubara frequencies $(2k+1)\pi T$, where $k$ is an 
integer.  As is customary in chiral random matrix models, we have 
retained only the lowest Matsubara frequency, $\pi T$.

It is conventional in $\chi$RMM to replace both the known form of the 
free Dirac operator as well as the gluon field contributions in the 
random matrix $W$.  The new feature of the present model is that the 
free Dirac operator is explicitly retained and $W$ is assumed to describe 
only gluonic contributions.  We choose the $N \times N$ matrix $W$ to 
be a random complex matrix, and its entries are drawn at random on a 
Gaussian distribution.  This choice of $W$ is motivated by the 
conjecture \cite{HV} that the correlators of QCD with three colours are 
given by the chiral Gaussian unitary ensemble.  (The Gaussian orthogonal 
and symplectic ensembles are expected to be relevant for the description 
of QCD with a smaller number of colours.)  Note that the form of the 
$W$-dependent part of $X$ in eqn.\,(\ref{Hmodel}) provides the most general 
antihermitian matrix which preserves the relation, $\gamma_5 X\gamma_5 = 
X^\dagger$, and thus ensures the hemiticity of $H$.  The only ``dynamical'' 
information in $W$ is its variance.  We shall consider this point below.
\vspace{3mm}

To see that this implementation of the GW relation yields a physically 
sensible non-doubled spectrum for the Dirac operator, it is useful to 
look at the free spectrum.  Turning off the gluon field, i.e., $W=0$,
the eigenvalues of the free Dirac operator can be found analytically.  
Diagonalizing $D$, one finds that the eigenvalues, $\lambda$, of $D$ are 
in one-to-one correspondence with the eigenvalues, 
$\xi_{j\pm}=1-a\Delta'_{jj}-ai\Delta_{jj}$, of the free Wilson 
operator $X$.  The result is simply $\lambda =(1-\xi/|\xi|)/a$. Evidently, 
only the phase information of 
$\xi$ is carried through to $\lambda$.  This information, however, is all 
that is needed in order to resolve the fermion doubling problem: For small 
momenta, the Wilson term makes a negligible contribution to the real part of 
$\xi$. The corresponding $\lambda$ eigenvalues appear at small angles, 
$\theta$, on the circle $(1-e^{i\theta})/a$.  As the momentum increases, 
the Wilson term becomes increasingly important with the effect that the 
doubled states are pushed towards $\theta=\pm \pi$. Since the doubled states
are thus well separated from the small $\theta$ region of physical interest, 
this represents a sensible resolution of the doubling problem expected to 
remain valid when $W$ is reintroduced.  When $W$ is not equal to zero, 
$X$ is no longer normal and such analytic evaluation is not possible to our 
knowledge.  (Of course, $H = \gamma_5 X$ is hermitian for both interacting 
and non-interacting fermions.)
\vspace{3mm}

In general the analytical relation between the eigenvalues of $D$
and the eigenvalues of $X$, $\lambda =(1-\xi/|\xi|)/a$, holds 
if $X$ is normal and satisfies $\gamma_5 X \gamma_5=X^\dagger$. Thus, 
this relation also applies when there is no deterministic part in $X$, 
i.e., $\Delta=\Delta'=0$.  The eigenvalues of $D$ can then be expressed 
in terms of the eigenvalues, $i\omega_j$, of 
the matrix
\begin{equation}
  \left(\begin{array}{cc}
      0 & iW \\
      iW^\dagger & 0
    \end{array}\right)
\end{equation}
which has the usual form of chiral random matrices.  The eigenvalues 
$i \omega_j$ are simply mapped onto the circle $(1-e^{i\theta_j})/a$ 
according to $\cos\theta_j=1/\sqrt{1+\omega_j}$.  This is sufficient 
to ensure that both the microscopic spectral density and the various spectral 
correlators of $D$ are identical to those of the original $\chi$RMM.  
   
\vspace{2mm}

{\bf \large \noindent Applications and numerical results}
\vspace{3mm}

The simplest application of the random matrix model constructed in the 
preceding section is the study of the chiral condensate as a function of 
temperature.  As indicated by eqn.\,(\ref{16}), this requires us to 
consider the microscopic spectral density of $D$.  We wish to consider 
the case where chiral symmetry is spontaneously broken at $T=0$.  In 
ordinary random matrix models, the scale of the problem is set solely 
by the variance of the random matrix $W$.  Chiral symmetry is always 
spontaneously broken at $T=0$, and the critical temperature for its 
restoration scales strictly with the variance of $W$.  The situation is 
somewhat more complicated in the present case where $D$ contains both 
deterministic and random elements.  Here, the existence of a chiral 
condensate even at zero temperature is determined by a competition 
between low-energy free eigenvalues and the variance of the random matrix 
$W$.  The spectral properties of matrices which are the sum of deterministic 
and random parts have been considered in refs.\,\cite{WSW,GuhrVeide,Z-J} and 
\cite{TandT}.  This work shows that the resulting spectral correlators 
and the microscopic spectral density will be given exactly by the 
strict random matrix results provided only that the random part of the 
matrix is sufficiently strong.  We expect to find a similar result in the 
present case.  
   
The problem considered explicitly in \cite{WSW} was a matrix of the form 
\begin{equation}
  \left(\begin{array}{cc}
      0 & \Delta+W \\
      \Delta+W^\dagger & 0
    \end{array}\right)
  \label{naiveD}
\end{equation}
where $\Delta$ is a fixed diagonal matrix and the elements of $W$ are 
drawn at random on the Gaussian weight 
\begin{equation}
P(W) \sim \exp\{-N \Sigma^2 {\rm Tr}WW^\dagger\} \ .
\label{weightII}
\end{equation}
The parameter $\Sigma$ introduced here determines the variance of the 
distribution.  When the diagonal elements of $\Delta$ are all non-zero, 
the chiral condensate, $\Xi$, can be expressed as 
\begin{equation}
\Xi=\Sigma^2\bar{x} \ ,
\end{equation}
where $\bar{x}$ is the only real and positive solution of
\begin{equation}
\label{defx}
N\Sigma^2=\sum_{j} \, \frac{1}{\Delta^2_{jj}+\bar{x}^2} \ .
\end{equation}
In the absence of a solution to eqn.\,(\ref{defx}), the chiral condensate is 
zero.  Noting that the right side of eqn.\,(\ref{defx}) is a monotonically 
decreasing function of $\bar{x}$, it follows that a necessary condition for 
a non-zero chiral condensate in the presence of the deterministic matrix 
$\Delta$ is
\begin{equation}
N\Sigma^2 \leq \sum_{j} \, \frac{1}{\Delta^2_{jj}} \ .
\label{varcondition}
\end{equation}

Of course, neither the matrices $X$, $H$, or $D$ of the present model have 
the form of eqn.\,(\ref{naiveD}).  Nevertheless, we would like to use 
eqns.\,(\ref{defx}) and (\ref{varcondition}) to provide a rough estimate 
of the variance of $W$ required to ensure a reasonable critical temperature 
for chiral symmetry restoration.  These equations are naturally dominated 
by the lowest eigenvalues of $\Delta$.  Turning to the matrix $X$ defined 
by eqns.\,(\ref{Hmodel})--(\ref{defdeltap}), we see that these lowest 
states correspond to the ``undoubled'' fermion states.  Restricting 
our attention only to these states, it is reasonable to neglect the Wilson 
term, $\Delta'$.  This leaves us with a matrix having the form of 
eqn.\,(\ref{naiveD}).  

Having first determined the size, $N$, of the matrix and the lattice 
spacing, $a$, we pick the variance of $W$ in order to fix the estimated 
critical temperature at some value, $T_c$.  Thus,
\begin{equation}
N\Sigma^2=\sum_{j=1}^{N/8}\frac{1}{(\Delta_{jj}(T_c))^2} \ .
\end{equation}
Numerical simulations indicate that this procedure is sound in practice and 
that the resulting critical temperature is close to this estimated value.

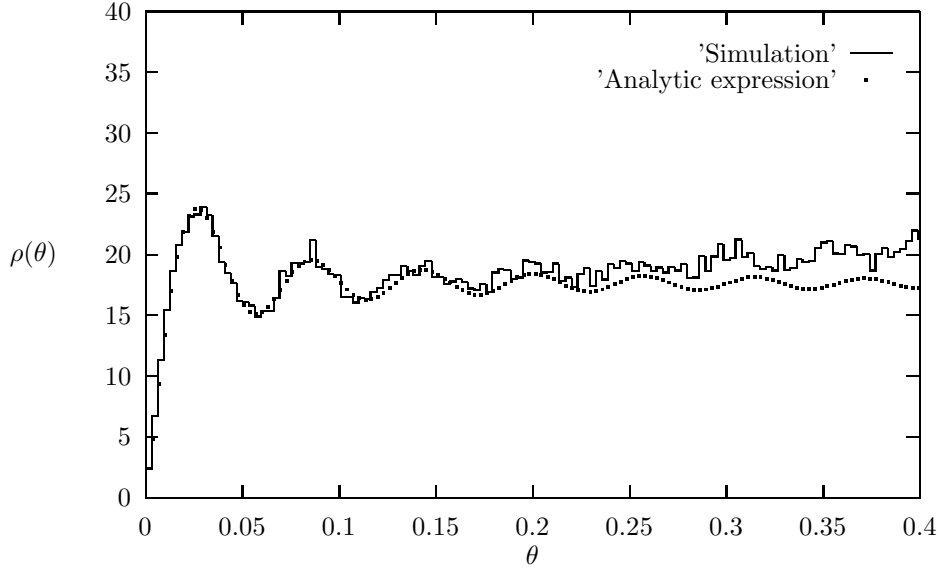
\begin{figure}[t]
\setlength{\unitlength}{0.240900pt}
\ifx\plotpoint\undefined\newsavebox{\plotpoint}\fi
\begin{picture}(1500,900)(0,0)
\font\gnuplot=cmr10 at 10pt
\gnuplot
\sbox{\plotpoint}{\rule[-0.200pt]{0.400pt}{0.400pt}}%
\put(220.0,113.0){\rule[-0.200pt]{292.934pt}{0.400pt}}
\put(220.0,113.0){\rule[-0.200pt]{0.400pt}{184.048pt}}
\put(220.0,113.0){\rule[-0.200pt]{4.818pt}{0.400pt}}
\put(198,113){\makebox(0,0)[r]{0}}
\put(1416.0,113.0){\rule[-0.200pt]{4.818pt}{0.400pt}}
\put(220.0,209.0){\rule[-0.200pt]{4.818pt}{0.400pt}}
\put(198,209){\makebox(0,0)[r]{5}}
\put(1416.0,209.0){\rule[-0.200pt]{4.818pt}{0.400pt}}
\put(220.0,304.0){\rule[-0.200pt]{4.818pt}{0.400pt}}
\put(198,304){\makebox(0,0)[r]{10}}
\put(1416.0,304.0){\rule[-0.200pt]{4.818pt}{0.400pt}}
\put(220.0,400.0){\rule[-0.200pt]{4.818pt}{0.400pt}}
\put(198,400){\makebox(0,0)[r]{15}}
\put(1416.0,400.0){\rule[-0.200pt]{4.818pt}{0.400pt}}
\put(220.0,495.0){\rule[-0.200pt]{4.818pt}{0.400pt}}
\put(198,495){\makebox(0,0)[r]{20}}
\put(1416.0,495.0){\rule[-0.200pt]{4.818pt}{0.400pt}}
\put(220.0,591.0){\rule[-0.200pt]{4.818pt}{0.400pt}}
\put(198,591){\makebox(0,0)[r]{25}}
\put(1416.0,591.0){\rule[-0.200pt]{4.818pt}{0.400pt}}
\put(220.0,686.0){\rule[-0.200pt]{4.818pt}{0.400pt}}
\put(198,686){\makebox(0,0)[r]{30}}
\put(1416.0,686.0){\rule[-0.200pt]{4.818pt}{0.400pt}}
\put(220.0,782.0){\rule[-0.200pt]{4.818pt}{0.400pt}}
\put(198,782){\makebox(0,0)[r]{35}}
\put(1416.0,782.0){\rule[-0.200pt]{4.818pt}{0.400pt}}
\put(220.0,877.0){\rule[-0.200pt]{4.818pt}{0.400pt}}
\put(198,877){\makebox(0,0)[r]{40}}
\put(1416.0,877.0){\rule[-0.200pt]{4.818pt}{0.400pt}}
\put(220.0,113.0){\rule[-0.200pt]{0.400pt}{4.818pt}}
\put(220,68){\makebox(0,0){0}}
\put(220.0,857.0){\rule[-0.200pt]{0.400pt}{4.818pt}}
\put(372.0,113.0){\rule[-0.200pt]{0.400pt}{4.818pt}}
\put(372,68){\makebox(0,0){0.05}}
\put(372.0,857.0){\rule[-0.200pt]{0.400pt}{4.818pt}}
\put(524.0,113.0){\rule[-0.200pt]{0.400pt}{4.818pt}}
\put(524,68){\makebox(0,0){0.1}}
\put(524.0,857.0){\rule[-0.200pt]{0.400pt}{4.818pt}}
\put(676.0,113.0){\rule[-0.200pt]{0.400pt}{4.818pt}}
\put(676,68){\makebox(0,0){0.15}}
\put(676.0,857.0){\rule[-0.200pt]{0.400pt}{4.818pt}}
\put(828.0,113.0){\rule[-0.200pt]{0.400pt}{4.818pt}}
\put(828,68){\makebox(0,0){0.2}}
\put(828.0,857.0){\rule[-0.200pt]{0.400pt}{4.818pt}}
\put(980.0,113.0){\rule[-0.200pt]{0.400pt}{4.818pt}}
\put(980,68){\makebox(0,0){0.25}}
\put(980.0,857.0){\rule[-0.200pt]{0.400pt}{4.818pt}}
\put(1132.0,113.0){\rule[-0.200pt]{0.400pt}{4.818pt}}
\put(1132,68){\makebox(0,0){0.3}}
\put(1132.0,857.0){\rule[-0.200pt]{0.400pt}{4.818pt}}
\put(1284.0,113.0){\rule[-0.200pt]{0.400pt}{4.818pt}}
\put(1284,68){\makebox(0,0){0.35}}
\put(1284.0,857.0){\rule[-0.200pt]{0.400pt}{4.818pt}}
\put(1436.0,113.0){\rule[-0.200pt]{0.400pt}{4.818pt}}
\put(1436,68){\makebox(0,0){0.4}}
\put(1436.0,857.0){\rule[-0.200pt]{0.400pt}{4.818pt}}
\put(220.0,113.0){\rule[-0.200pt]{292.934pt}{0.400pt}}
\put(1436.0,113.0){\rule[-0.200pt]{0.400pt}{184.048pt}}
\put(220.0,877.0){\rule[-0.200pt]{292.934pt}{0.400pt}}
\put(45,495){\makebox(0,0){$\rho(\theta)$}}
\put(828,23){\makebox(0,0){$\theta$}}
\put(220.0,113.0){\rule[-0.200pt]{0.400pt}{184.048pt}}
\put(1306,812){\makebox(0,0)[r]{'Simulation'}}
\put(1328.0,812.0){\rule[-0.200pt]{15.899pt}{0.400pt}}
\put(220,159){\usebox{\plotpoint}}
\put(220,159){\usebox{\plotpoint}}
\put(220.0,159.0){\rule[-0.200pt]{2.409pt}{0.400pt}}
\put(230.0,159.0){\rule[-0.200pt]{0.400pt}{19.995pt}}
\put(230.0,242.0){\rule[-0.200pt]{2.168pt}{0.400pt}}
\put(239.0,242.0){\rule[-0.200pt]{0.400pt}{20.958pt}}
\put(239.0,329.0){\rule[-0.200pt]{2.409pt}{0.400pt}}
\put(249.0,329.0){\rule[-0.200pt]{0.400pt}{19.031pt}}
\put(249.0,408.0){\rule[-0.200pt]{2.168pt}{0.400pt}}
\put(258.0,408.0){\rule[-0.200pt]{0.400pt}{14.695pt}}
\put(258.0,469.0){\rule[-0.200pt]{2.409pt}{0.400pt}}
\put(268.0,469.0){\rule[-0.200pt]{0.400pt}{9.877pt}}
\put(268.0,510.0){\rule[-0.200pt]{2.168pt}{0.400pt}}
\put(277.0,510.0){\rule[-0.200pt]{0.400pt}{5.059pt}}
\put(277.0,531.0){\rule[-0.200pt]{2.409pt}{0.400pt}}
\put(287.0,531.0){\rule[-0.200pt]{0.400pt}{5.782pt}}
\put(287.0,555.0){\rule[-0.200pt]{2.168pt}{0.400pt}}
\put(296.0,555.0){\rule[-0.200pt]{0.400pt}{0.723pt}}
\put(296.0,558.0){\rule[-0.200pt]{2.409pt}{0.400pt}}
\put(306.0,558.0){\rule[-0.200pt]{0.400pt}{2.891pt}}
\put(306.0,570.0){\rule[-0.200pt]{2.409pt}{0.400pt}}
\put(316.0,557.0){\rule[-0.200pt]{0.400pt}{3.132pt}}
\put(316.0,557.0){\rule[-0.200pt]{2.168pt}{0.400pt}}
\put(325.0,524.0){\rule[-0.200pt]{0.400pt}{7.950pt}}
\put(325.0,524.0){\rule[-0.200pt]{2.409pt}{0.400pt}}
\put(335.0,484.0){\rule[-0.200pt]{0.400pt}{9.636pt}}
\put(335.0,484.0){\rule[-0.200pt]{2.168pt}{0.400pt}}
\put(344.0,466.0){\rule[-0.200pt]{0.400pt}{4.336pt}}
\put(344.0,466.0){\rule[-0.200pt]{2.409pt}{0.400pt}}
\put(354.0,450.0){\rule[-0.200pt]{0.400pt}{3.854pt}}
\put(354.0,450.0){\rule[-0.200pt]{2.168pt}{0.400pt}}
\put(363.0,422.0){\rule[-0.200pt]{0.400pt}{6.745pt}}
\put(363.0,422.0){\rule[-0.200pt]{2.409pt}{0.400pt}}
\put(373.0,421.0){\usebox{\plotpoint}}
\put(373.0,421.0){\rule[-0.200pt]{2.168pt}{0.400pt}}
\put(382.0,415.0){\rule[-0.200pt]{0.400pt}{1.445pt}}
\put(382.0,415.0){\rule[-0.200pt]{2.409pt}{0.400pt}}
\put(392.0,398.0){\rule[-0.200pt]{0.400pt}{4.095pt}}
\put(392.0,398.0){\rule[-0.200pt]{2.168pt}{0.400pt}}
\put(401.0,398.0){\rule[-0.200pt]{0.400pt}{2.168pt}}
\put(401.0,407.0){\rule[-0.200pt]{4.818pt}{0.400pt}}
\put(421.0,407.0){\rule[-0.200pt]{0.400pt}{4.818pt}}
\put(421.0,427.0){\rule[-0.200pt]{2.168pt}{0.400pt}}
\put(430.0,427.0){\rule[-0.200pt]{0.400pt}{10.118pt}}
\put(430.0,469.0){\rule[-0.200pt]{2.409pt}{0.400pt}}
\put(440.0,462.0){\rule[-0.200pt]{0.400pt}{1.686pt}}
\put(440.0,462.0){\rule[-0.200pt]{2.168pt}{0.400pt}}
\put(449.0,462.0){\rule[-0.200pt]{0.400pt}{4.818pt}}
\put(449.0,482.0){\rule[-0.200pt]{2.409pt}{0.400pt}}
\put(459.0,481.0){\usebox{\plotpoint}}
\put(459.0,481.0){\rule[-0.200pt]{2.168pt}{0.400pt}}
\put(468.0,481.0){\rule[-0.200pt]{0.400pt}{0.723pt}}
\put(468.0,484.0){\rule[-0.200pt]{2.409pt}{0.400pt}}
\put(478.0,484.0){\rule[-0.200pt]{0.400pt}{8.191pt}}
\put(478.0,518.0){\rule[-0.200pt]{2.168pt}{0.400pt}}
\put(487.0,476.0){\rule[-0.200pt]{0.400pt}{10.118pt}}
\put(487.0,476.0){\rule[-0.200pt]{2.409pt}{0.400pt}}
\put(497.0,473.0){\rule[-0.200pt]{0.400pt}{0.723pt}}
\put(497.0,473.0){\rule[-0.200pt]{2.409pt}{0.400pt}}
\put(507.0,465.0){\rule[-0.200pt]{0.400pt}{1.927pt}}
\put(507.0,465.0){\rule[-0.200pt]{2.168pt}{0.400pt}}
\put(516.0,463.0){\rule[-0.200pt]{0.400pt}{0.482pt}}
\put(516.0,463.0){\rule[-0.200pt]{2.409pt}{0.400pt}}
\put(526.0,429.0){\rule[-0.200pt]{0.400pt}{8.191pt}}
\put(526.0,429.0){\rule[-0.200pt]{4.577pt}{0.400pt}}
\put(545.0,420.0){\rule[-0.200pt]{0.400pt}{2.168pt}}
\put(545.0,420.0){\rule[-0.200pt]{2.168pt}{0.400pt}}
\put(554.0,420.0){\rule[-0.200pt]{0.400pt}{1.686pt}}
\put(554.0,427.0){\rule[-0.200pt]{2.409pt}{0.400pt}}
\put(564.0,425.0){\rule[-0.200pt]{0.400pt}{0.482pt}}
\put(564.0,425.0){\rule[-0.200pt]{2.168pt}{0.400pt}}
\put(573.0,425.0){\rule[-0.200pt]{0.400pt}{2.650pt}}
\put(573.0,436.0){\rule[-0.200pt]{2.409pt}{0.400pt}}
\put(583.0,436.0){\rule[-0.200pt]{0.400pt}{1.686pt}}
\put(583.0,443.0){\rule[-0.200pt]{2.168pt}{0.400pt}}
\put(592.0,443.0){\rule[-0.200pt]{0.400pt}{2.891pt}}
\put(592.0,455.0){\rule[-0.200pt]{2.409pt}{0.400pt}}
\put(602.0,455.0){\rule[-0.200pt]{0.400pt}{1.927pt}}
\put(602.0,463.0){\rule[-0.200pt]{4.577pt}{0.400pt}}
\put(621.0,463.0){\rule[-0.200pt]{0.400pt}{3.373pt}}
\put(621.0,477.0){\rule[-0.200pt]{2.409pt}{0.400pt}}
\put(631.0,469.0){\rule[-0.200pt]{0.400pt}{1.927pt}}
\put(631.0,469.0){\rule[-0.200pt]{2.168pt}{0.400pt}}
\put(640.0,463.0){\rule[-0.200pt]{0.400pt}{1.445pt}}
\put(640.0,463.0){\rule[-0.200pt]{2.409pt}{0.400pt}}
\put(650.0,463.0){\rule[-0.200pt]{0.400pt}{3.373pt}}
\put(650.0,477.0){\rule[-0.200pt]{2.168pt}{0.400pt}}
\put(659.0,477.0){\rule[-0.200pt]{0.400pt}{1.927pt}}
\put(659.0,485.0){\rule[-0.200pt]{2.409pt}{0.400pt}}
\put(669.0,460.0){\rule[-0.200pt]{0.400pt}{6.022pt}}
\put(669.0,460.0){\rule[-0.200pt]{4.577pt}{0.400pt}}
\put(688.0,449.0){\rule[-0.200pt]{0.400pt}{2.650pt}}
\put(688.0,449.0){\rule[-0.200pt]{2.409pt}{0.400pt}}
\put(698.0,449.0){\rule[-0.200pt]{0.400pt}{0.964pt}}
\put(698.0,453.0){\rule[-0.200pt]{2.168pt}{0.400pt}}
\put(707.0,453.0){\rule[-0.200pt]{0.400pt}{0.964pt}}
\put(707.0,457.0){\rule[-0.200pt]{2.409pt}{0.400pt}}
\put(717.0,452.0){\rule[-0.200pt]{0.400pt}{1.204pt}}
\put(717.0,452.0){\rule[-0.200pt]{2.168pt}{0.400pt}}
\put(726.0,442.0){\rule[-0.200pt]{0.400pt}{2.409pt}}
\put(726.0,442.0){\rule[-0.200pt]{2.409pt}{0.400pt}}
\put(736.0,439.0){\rule[-0.200pt]{0.400pt}{0.723pt}}
\put(736.0,439.0){\rule[-0.200pt]{2.168pt}{0.400pt}}
\put(745.0,439.0){\rule[-0.200pt]{0.400pt}{2.409pt}}
\put(745.0,449.0){\rule[-0.200pt]{2.409pt}{0.400pt}}
\put(755.0,436.0){\rule[-0.200pt]{0.400pt}{3.132pt}}
\put(755.0,436.0){\rule[-0.200pt]{2.168pt}{0.400pt}}
\put(764.0,436.0){\rule[-0.200pt]{0.400pt}{7.468pt}}
\put(764.0,467.0){\rule[-0.200pt]{2.409pt}{0.400pt}}
\put(774.0,467.0){\rule[-0.200pt]{0.400pt}{1.204pt}}
\put(774.0,472.0){\rule[-0.200pt]{2.168pt}{0.400pt}}
\put(783.0,471.0){\usebox{\plotpoint}}
\put(783.0,471.0){\rule[-0.200pt]{2.409pt}{0.400pt}}
\put(793.0,457.0){\rule[-0.200pt]{0.400pt}{3.373pt}}
\put(793.0,457.0){\rule[-0.200pt]{2.409pt}{0.400pt}}
\put(803.0,457.0){\rule[-0.200pt]{0.400pt}{0.482pt}}
\put(803.0,459.0){\rule[-0.200pt]{2.168pt}{0.400pt}}
\put(812.0,459.0){\rule[-0.200pt]{0.400pt}{6.504pt}}
\put(812.0,486.0){\rule[-0.200pt]{2.409pt}{0.400pt}}
\put(822.0,484.0){\rule[-0.200pt]{0.400pt}{0.482pt}}
\put(822.0,484.0){\rule[-0.200pt]{2.168pt}{0.400pt}}
\put(831.0,481.0){\rule[-0.200pt]{0.400pt}{0.723pt}}
\put(831.0,481.0){\rule[-0.200pt]{2.409pt}{0.400pt}}
\put(841.0,467.0){\rule[-0.200pt]{0.400pt}{3.373pt}}
\put(841.0,467.0){\rule[-0.200pt]{2.168pt}{0.400pt}}
\put(850.0,467.0){\usebox{\plotpoint}}
\put(850.0,468.0){\rule[-0.200pt]{2.409pt}{0.400pt}}
\put(860.0,468.0){\rule[-0.200pt]{0.400pt}{3.132pt}}
\put(860.0,481.0){\rule[-0.200pt]{2.168pt}{0.400pt}}
\put(869.0,454.0){\rule[-0.200pt]{0.400pt}{6.504pt}}
\put(869.0,454.0){\rule[-0.200pt]{2.409pt}{0.400pt}}
\put(879.0,454.0){\rule[-0.200pt]{0.400pt}{1.204pt}}
\put(879.0,459.0){\rule[-0.200pt]{2.409pt}{0.400pt}}
\put(889.0,444.0){\rule[-0.200pt]{0.400pt}{3.613pt}}
\put(889.0,444.0){\rule[-0.200pt]{2.168pt}{0.400pt}}
\put(898.0,444.0){\rule[-0.200pt]{0.400pt}{5.059pt}}
\put(898.0,465.0){\rule[-0.200pt]{2.409pt}{0.400pt}}
\put(908.0,465.0){\rule[-0.200pt]{0.400pt}{2.168pt}}
\put(908.0,474.0){\rule[-0.200pt]{2.168pt}{0.400pt}}
\put(917.0,446.0){\rule[-0.200pt]{0.400pt}{6.745pt}}
\put(917.0,446.0){\rule[-0.200pt]{2.409pt}{0.400pt}}
\put(927.0,446.0){\rule[-0.200pt]{0.400pt}{5.541pt}}
\put(927.0,469.0){\rule[-0.200pt]{2.168pt}{0.400pt}}
\put(936.0,456.0){\rule[-0.200pt]{0.400pt}{3.132pt}}
\put(936.0,456.0){\rule[-0.200pt]{2.409pt}{0.400pt}}
\put(946.0,456.0){\rule[-0.200pt]{0.400pt}{5.782pt}}
\put(946.0,480.0){\rule[-0.200pt]{2.168pt}{0.400pt}}
\put(955.0,474.0){\rule[-0.200pt]{0.400pt}{1.445pt}}
\put(955.0,474.0){\rule[-0.200pt]{2.409pt}{0.400pt}}
\put(965.0,474.0){\rule[-0.200pt]{0.400pt}{2.650pt}}
\put(965.0,485.0){\rule[-0.200pt]{2.168pt}{0.400pt}}
\put(974.0,481.0){\rule[-0.200pt]{0.400pt}{0.964pt}}
\put(974.0,481.0){\rule[-0.200pt]{2.409pt}{0.400pt}}
\put(984.0,469.0){\rule[-0.200pt]{0.400pt}{2.891pt}}
\put(984.0,469.0){\rule[-0.200pt]{2.409pt}{0.400pt}}
\put(994.0,469.0){\rule[-0.200pt]{0.400pt}{3.613pt}}
\put(994.0,484.0){\rule[-0.200pt]{2.168pt}{0.400pt}}
\put(1003.0,478.0){\rule[-0.200pt]{0.400pt}{1.445pt}}
\put(1003.0,478.0){\rule[-0.200pt]{2.409pt}{0.400pt}}
\put(1013.0,469.0){\rule[-0.200pt]{0.400pt}{2.168pt}}
\put(1013.0,469.0){\rule[-0.200pt]{2.168pt}{0.400pt}}
\put(1022.0,469.0){\rule[-0.200pt]{0.400pt}{1.445pt}}
\put(1022.0,475.0){\rule[-0.200pt]{2.409pt}{0.400pt}}
\put(1032.0,467.0){\rule[-0.200pt]{0.400pt}{1.927pt}}
\put(1032.0,467.0){\rule[-0.200pt]{2.168pt}{0.400pt}}
\put(1041.0,467.0){\rule[-0.200pt]{0.400pt}{2.168pt}}
\put(1041.0,476.0){\rule[-0.200pt]{2.409pt}{0.400pt}}
\put(1051.0,470.0){\rule[-0.200pt]{0.400pt}{1.445pt}}
\put(1051.0,470.0){\rule[-0.200pt]{2.168pt}{0.400pt}}
\put(1060.0,470.0){\rule[-0.200pt]{0.400pt}{2.409pt}}
\put(1060.0,480.0){\rule[-0.200pt]{2.409pt}{0.400pt}}
\put(1070.0,458.0){\rule[-0.200pt]{0.400pt}{5.300pt}}
\put(1070.0,458.0){\rule[-0.200pt]{2.409pt}{0.400pt}}
\put(1080.0,458.0){\usebox{\plotpoint}}
\put(1080.0,459.0){\rule[-0.200pt]{2.168pt}{0.400pt}}
\put(1089.0,459.0){\rule[-0.200pt]{0.400pt}{8.191pt}}
\put(1089.0,493.0){\rule[-0.200pt]{2.409pt}{0.400pt}}
\put(1099.0,469.0){\rule[-0.200pt]{0.400pt}{5.782pt}}
\put(1099.0,469.0){\rule[-0.200pt]{2.168pt}{0.400pt}}
\put(1108.0,469.0){\rule[-0.200pt]{0.400pt}{5.541pt}}
\put(1108.0,492.0){\rule[-0.200pt]{2.409pt}{0.400pt}}
\put(1118.0,492.0){\rule[-0.200pt]{0.400pt}{4.818pt}}
\put(1118.0,512.0){\rule[-0.200pt]{2.168pt}{0.400pt}}
\put(1127.0,487.0){\rule[-0.200pt]{0.400pt}{6.022pt}}
\put(1127.0,487.0){\rule[-0.200pt]{2.409pt}{0.400pt}}
\put(1137.0,487.0){\rule[-0.200pt]{0.400pt}{0.723pt}}
\put(1137.0,490.0){\rule[-0.200pt]{2.168pt}{0.400pt}}
\put(1146.0,490.0){\rule[-0.200pt]{0.400pt}{6.986pt}}
\put(1146.0,519.0){\rule[-0.200pt]{2.409pt}{0.400pt}}
\put(1156.0,491.0){\rule[-0.200pt]{0.400pt}{6.745pt}}
\put(1156.0,491.0){\rule[-0.200pt]{2.168pt}{0.400pt}}
\put(1165.0,491.0){\rule[-0.200pt]{0.400pt}{1.686pt}}
\put(1165.0,498.0){\rule[-0.200pt]{2.409pt}{0.400pt}}
\put(1175.0,477.0){\rule[-0.200pt]{0.400pt}{5.059pt}}
\put(1175.0,477.0){\rule[-0.200pt]{2.409pt}{0.400pt}}
\put(1185.0,472.0){\rule[-0.200pt]{0.400pt}{1.204pt}}
\put(1185.0,472.0){\rule[-0.200pt]{2.168pt}{0.400pt}}
\put(1194.0,472.0){\rule[-0.200pt]{0.400pt}{0.964pt}}
\put(1194.0,476.0){\rule[-0.200pt]{2.409pt}{0.400pt}}
\put(1204.0,470.0){\rule[-0.200pt]{0.400pt}{1.445pt}}
\put(1204.0,470.0){\rule[-0.200pt]{2.168pt}{0.400pt}}
\put(1213.0,470.0){\rule[-0.200pt]{0.400pt}{5.782pt}}
\put(1213.0,494.0){\rule[-0.200pt]{2.409pt}{0.400pt}}
\put(1223.0,471.0){\rule[-0.200pt]{0.400pt}{5.541pt}}
\put(1223.0,471.0){\rule[-0.200pt]{2.168pt}{0.400pt}}
\put(1232.0,471.0){\rule[-0.200pt]{0.400pt}{0.723pt}}
\put(1232.0,474.0){\rule[-0.200pt]{2.409pt}{0.400pt}}
\put(1242.0,474.0){\rule[-0.200pt]{0.400pt}{3.373pt}}
\put(1242.0,488.0){\rule[-0.200pt]{2.168pt}{0.400pt}}
\put(1251.0,484.0){\rule[-0.200pt]{0.400pt}{0.964pt}}
\put(1251.0,484.0){\rule[-0.200pt]{2.409pt}{0.400pt}}
\put(1261.0,484.0){\usebox{\plotpoint}}
\put(1261.0,485.0){\rule[-0.200pt]{2.409pt}{0.400pt}}
\put(1271.0,485.0){\rule[-0.200pt]{0.400pt}{5.059pt}}
\put(1271.0,506.0){\rule[-0.200pt]{2.168pt}{0.400pt}}
\put(1280.0,506.0){\rule[-0.200pt]{0.400pt}{1.927pt}}
\put(1280.0,514.0){\rule[-0.200pt]{2.409pt}{0.400pt}}
\put(1290.0,514.0){\rule[-0.200pt]{0.400pt}{0.482pt}}
\put(1290.0,516.0){\rule[-0.200pt]{2.168pt}{0.400pt}}
\put(1299.0,497.0){\rule[-0.200pt]{0.400pt}{4.577pt}}
\put(1299.0,497.0){\rule[-0.200pt]{2.409pt}{0.400pt}}
\put(1309.0,496.0){\usebox{\plotpoint}}
\put(1309.0,496.0){\rule[-0.200pt]{2.168pt}{0.400pt}}
\put(1318.0,496.0){\rule[-0.200pt]{0.400pt}{3.373pt}}
\put(1318.0,510.0){\rule[-0.200pt]{2.409pt}{0.400pt}}
\put(1328.0,508.0){\rule[-0.200pt]{0.400pt}{0.482pt}}
\put(1328.0,508.0){\rule[-0.200pt]{2.168pt}{0.400pt}}
\put(1337.0,495.0){\rule[-0.200pt]{0.400pt}{3.132pt}}
\put(1337.0,495.0){\rule[-0.200pt]{2.409pt}{0.400pt}}
\put(1347.0,495.0){\usebox{\plotpoint}}
\put(1347.0,496.0){\rule[-0.200pt]{2.409pt}{0.400pt}}
\put(1357.0,470.0){\rule[-0.200pt]{0.400pt}{6.263pt}}
\put(1357.0,470.0){\rule[-0.200pt]{2.168pt}{0.400pt}}
\put(1366.0,470.0){\rule[-0.200pt]{0.400pt}{6.263pt}}
\put(1366.0,496.0){\rule[-0.200pt]{2.409pt}{0.400pt}}
\put(1376.0,496.0){\rule[-0.200pt]{0.400pt}{2.168pt}}
\put(1376.0,505.0){\rule[-0.200pt]{2.168pt}{0.400pt}}
\put(1385.0,491.0){\rule[-0.200pt]{0.400pt}{3.373pt}}
\put(1385.0,491.0){\rule[-0.200pt]{2.409pt}{0.400pt}}
\put(1395.0,491.0){\rule[-0.200pt]{0.400pt}{1.927pt}}
\put(1395.0,499.0){\rule[-0.200pt]{2.168pt}{0.400pt}}
\put(1404.0,499.0){\rule[-0.200pt]{0.400pt}{1.204pt}}
\put(1404.0,504.0){\rule[-0.200pt]{2.409pt}{0.400pt}}
\put(1414.0,504.0){\rule[-0.200pt]{0.400pt}{3.132pt}}
\put(1414.0,517.0){\rule[-0.200pt]{2.168pt}{0.400pt}}
\put(1423.0,517.0){\rule[-0.200pt]{0.400pt}{3.854pt}}
\put(1423.0,533.0){\rule[-0.200pt]{2.409pt}{0.400pt}}
\put(1433.0,520.0){\rule[-0.200pt]{0.400pt}{3.132pt}}
\put(1433.0,520.0){\rule[-0.200pt]{0.723pt}{0.400pt}}
\put(1306,767){\makebox(0,0)[r]{'Analytic expression'}}
\put(1350,767){\rule{1pt}{1pt}}
\put(230,203){\rule{1pt}{1pt}}
\put(239,289){\rule{1pt}{1pt}}
\put(249,367){\rule{1pt}{1pt}}
\put(258,435){\rule{1pt}{1pt}}
\put(268,489){\rule{1pt}{1pt}}
\put(277,529){\rule{1pt}{1pt}}
\put(287,554){\rule{1pt}{1pt}}
\put(296,565){\rule{1pt}{1pt}}
\put(306,563){\rule{1pt}{1pt}}
\put(315,550){\rule{1pt}{1pt}}
\put(325,529){\rule{1pt}{1pt}}
\put(335,504){\rule{1pt}{1pt}}
\put(344,477){\rule{1pt}{1pt}}
\put(354,452){\rule{1pt}{1pt}}
\put(363,430){\rule{1pt}{1pt}}
\put(373,413){\rule{1pt}{1pt}}
\put(382,403){\rule{1pt}{1pt}}
\put(392,399){\rule{1pt}{1pt}}
\put(401,402){\rule{1pt}{1pt}}
\put(411,410){\rule{1pt}{1pt}}
\put(420,423){\rule{1pt}{1pt}}
\put(430,437){\rule{1pt}{1pt}}
\put(440,451){\rule{1pt}{1pt}}
\put(449,465){\rule{1pt}{1pt}}
\put(459,475){\rule{1pt}{1pt}}
\put(468,482){\rule{1pt}{1pt}}
\put(478,484){\rule{1pt}{1pt}}
\put(487,483){\rule{1pt}{1pt}}
\put(497,477){\rule{1pt}{1pt}}
\put(506,469){\rule{1pt}{1pt}}
\put(516,459){\rule{1pt}{1pt}}
\put(525,448){\rule{1pt}{1pt}}
\put(535,438){\rule{1pt}{1pt}}
\put(545,430){\rule{1pt}{1pt}}
\put(554,425){\rule{1pt}{1pt}}
\put(564,422){\rule{1pt}{1pt}}
\put(573,423){\rule{1pt}{1pt}}
\put(583,426){\rule{1pt}{1pt}}
\put(592,432){\rule{1pt}{1pt}}
\put(602,440){\rule{1pt}{1pt}}
\put(611,448){\rule{1pt}{1pt}}
\put(621,456){\rule{1pt}{1pt}}
\put(630,462){\rule{1pt}{1pt}}
\put(640,467){\rule{1pt}{1pt}}
\put(650,469){\rule{1pt}{1pt}}
\put(659,469){\rule{1pt}{1pt}}
\put(669,466){\rule{1pt}{1pt}}
\put(678,461){\rule{1pt}{1pt}}
\put(688,455){\rule{1pt}{1pt}}
\put(697,449){\rule{1pt}{1pt}}
\put(707,442){\rule{1pt}{1pt}}
\put(716,437){\rule{1pt}{1pt}}
\put(726,433){\rule{1pt}{1pt}}
\put(735,430){\rule{1pt}{1pt}}
\put(745,430){\rule{1pt}{1pt}}
\put(755,432){\rule{1pt}{1pt}}
\put(764,436){\rule{1pt}{1pt}}
\put(774,441){\rule{1pt}{1pt}}
\put(783,447){\rule{1pt}{1pt}}
\put(793,452){\rule{1pt}{1pt}}
\put(802,457){\rule{1pt}{1pt}}
\put(812,461){\rule{1pt}{1pt}}
\put(821,463){\rule{1pt}{1pt}}
\put(831,463){\rule{1pt}{1pt}}
\put(840,461){\rule{1pt}{1pt}}
\put(850,458){\rule{1pt}{1pt}}
\put(860,454){\rule{1pt}{1pt}}
\put(869,449){\rule{1pt}{1pt}}
\put(879,444){\rule{1pt}{1pt}}
\put(888,440){\rule{1pt}{1pt}}
\put(898,437){\rule{1pt}{1pt}}
\put(907,435){\rule{1pt}{1pt}}
\put(917,434){\rule{1pt}{1pt}}
\put(926,436){\rule{1pt}{1pt}}
\put(936,438){\rule{1pt}{1pt}}
\put(945,442){\rule{1pt}{1pt}}
\put(955,446){\rule{1pt}{1pt}}
\put(965,450){\rule{1pt}{1pt}}
\put(974,454){\rule{1pt}{1pt}}
\put(984,457){\rule{1pt}{1pt}}
\put(993,459){\rule{1pt}{1pt}}
\put(1003,459){\rule{1pt}{1pt}}
\put(1012,458){\rule{1pt}{1pt}}
\put(1022,456){\rule{1pt}{1pt}}
\put(1031,453){\rule{1pt}{1pt}}
\put(1041,449){\rule{1pt}{1pt}}
\put(1050,445){\rule{1pt}{1pt}}
\put(1060,442){\rule{1pt}{1pt}}
\put(1070,439){\rule{1pt}{1pt}}
\put(1079,437){\rule{1pt}{1pt}}
\put(1089,437){\rule{1pt}{1pt}}
\put(1098,438){\rule{1pt}{1pt}}
\put(1108,440){\rule{1pt}{1pt}}
\put(1117,442){\rule{1pt}{1pt}}
\put(1127,446){\rule{1pt}{1pt}}
\put(1136,449){\rule{1pt}{1pt}}
\put(1146,452){\rule{1pt}{1pt}}
\put(1155,455){\rule{1pt}{1pt}}
\put(1165,457){\rule{1pt}{1pt}}
\put(1175,457){\rule{1pt}{1pt}}
\put(1184,457){\rule{1pt}{1pt}}
\put(1194,455){\rule{1pt}{1pt}}
\put(1203,453){\rule{1pt}{1pt}}
\put(1213,450){\rule{1pt}{1pt}}
\put(1222,446){\rule{1pt}{1pt}}
\put(1232,443){\rule{1pt}{1pt}}
\put(1241,441){\rule{1pt}{1pt}}
\put(1251,439){\rule{1pt}{1pt}}
\put(1260,439){\rule{1pt}{1pt}}
\put(1270,439){\rule{1pt}{1pt}}
\put(1280,440){\rule{1pt}{1pt}}
\put(1289,443){\rule{1pt}{1pt}}
\put(1299,445){\rule{1pt}{1pt}}
\put(1308,448){\rule{1pt}{1pt}}
\put(1318,451){\rule{1pt}{1pt}}
\put(1327,453){\rule{1pt}{1pt}}
\put(1337,455){\rule{1pt}{1pt}}
\put(1346,456){\rule{1pt}{1pt}}
\put(1356,455){\rule{1pt}{1pt}}
\put(1365,454){\rule{1pt}{1pt}}
\put(1375,452){\rule{1pt}{1pt}}
\put(1385,450){\rule{1pt}{1pt}}
\put(1394,447){\rule{1pt}{1pt}}
\put(1404,444){\rule{1pt}{1pt}}
\put(1413,442){\rule{1pt}{1pt}}
\put(1423,441){\rule{1pt}{1pt}}
\put(1432,440){\rule{1pt}{1pt}}
\end{picture}
\caption{Universal behaviour of the eigenvalue density for the Dirac operator 
of the local $\chi$RMM. The histogram shows the microscopic level density at 
zero temperature while the smooth dotted curve is the analytic Bessel function 
expression suitably scaled. This simulation was performed for a $24\times24$ 
lattice with a lattice spacing of 1/4 fm and a critical temperature of 
approximately 150 MeV.}
\end{figure}

As an initial application of this local $\chi$RMM, we have performed a 
numerical study of the microscopic level density.  As indicated above, 
we anticipate that the microscopic spectral density for our model will 
be identical to the usual $\chi$RMM result, which can be expressed in 
terms of the first two Bessel functions \cite{VerbaarZahed} and \cite{B-BI} 
as: 
\begin{equation}
\rho(\lambda)=2N\Xi(N\Xi\lambda)[J_0^2(2N\lambda\Xi)+J_1^2(2N\lambda\Xi)] \ .
\label{bessel}
\end{equation}
The analogous result for the Gaussian symplectic ensemble has been 
shown to be in excellent agreement with the results of QCD lattice simulations 
at low temperature.  Eqn.\,({\ref{bessel}) applies whenever chiral symmetry 
is spontaneously broken, and the magnitude of the chiral condensate is 
directly related to the asymptotic value of the microscopic spectral density.
As anticipated, numerical results obtained for a variety of temperatures 
below $T_c$ are in agreement with eqn.\,(\ref{bessel}).  Results for 
the $\rho$ at $T=0$ are shown in the figure.  The density is plotted as 
a function of $\theta$ with $\lambda = (1 - \exp[i \theta])/a$.  The 
disagreement between analytic and numerical results seen for larger values 
of $\theta$ is a consequence of both finite size effects and, more 
importantly, the limited ability of the random matrix, $W$, to mix free 
quark states of very different momenta.

\newpage

{\bf \large \noindent Conclusions and discussion}
\vspace{3mm}

We have suggested a simple and well-defined procedure for modifying the 
usual Wilson Dirac operator in order to satisfy the Ginsparg-Wilson 
relation.  The result is a lattice Dirac operator which both solves the 
fermion doubling problem and retains a lattice chiral symmetry for all 
lattice spacings and sizes.  As noted, this lattice chiral symmetry 
becomes genuine chiral symmetry in the continuum limit.  While we 
have used this procedure to construct a local chiral random matrix 
model, we emphasise that it is more generally applicable.  (For example, 
the procedure outlined here would be suitable for QCD lattice simulations.)  
As is common in $\chi$RMM, we have introduced temperature dependence 
through the lowest Matsubara frequency.  We have studied the microscopic 
level density of the resulting Dirac operator numerically and confirmed 
the anticipated agreement with the universal behaviour found empirically 
in QCD lattice data and analytically in other chiral random matrix models.  
This allows us to extract the continuum limit (i.e., $N \to \infty$) of 
the lattice chiral condensate from finite-size simulations.  The existence 
of this lattice chiral symmetry for every $a$ leads us to anticipate that 
the limit $a \to 0$, in which lattice chiral symmetry becomes genuine 
chiral symmetry, will be relatively smooth.  

This local $\chi$RMM represents an appealing intermediate step between 
pure random matrix models and real QCD lattice simulations.  There appear 
to be a variety of interesting applications of local $\chi$RMM.  The 
presence of the free Dirac operator (in a form free of fermion doubling 
problems) makes it sensible to consider a number of local properties 
which are genuinely not accessible to ordinary random matrix theory.  
As noted, it is possible to explore the interplay between random and 
deterministic elements in this model and to study the range of applicability  
of random matrix techniques to lattice simulations.  It is now possible to 
search for quasi-universal behaviour in correlators and to address questions 
regarding localisation.  While such studies will initially be numerical, 
they are far simpler than full QCD lattice simulations and may well 
have interesting insights to offer.

\vspace{3mm}

{\bf \large \noindent Acknowledgements:} We gratefully acknowledge 
useful discussions with Kari Rummukainen and clarifying correspondence 
with Martin L\"usher and Herbert Neuberger.

\end{document}